\documentclass[aps,preprint]{revtex4}
\usepackage{graphicx,amssymb,physics}
\begin{document}
\setcounter{page}{1}

\title{Quantum Bit Behavior of Pinned Fluxes on Volume Defects \\in a Superconductor }
\author{H. B. \surname{Lee}}
\email{superpig@pusan.ac.kr}
\thanks{Fax: +82-51-513-7664}
\author{G. C. \surname{Kim}}
\author{Byeong-Joo \surname{Kim}}
\author{Young Jin \surname{Sohn}}
\author{Y. C. \surname{Kim}}
\affiliation{Department of Physics, Pusan National University, Busan 46241, Korea}

\begin{abstract}
We studied a qubit based on  flux-pinning effects in $\Delta$H=$\Delta$B region of a superconductor. When volume defects are many enough in a superconductor,  $\Delta$H=$\Delta$B  region on M-H curve is formed, which is the region that  increased applied magnetic field ($\Delta$H) is the same as increasing  magnetic induction ($\Delta$B).  Magnetization (M) is constant in the region by 4$\pi$M = B - H. Here we show that the behavior of fluxes in $\Delta$H=$\Delta$B  region can be a candidate of qubit.  Pinned fluxes on volume defects would move as a bundle in the region by repeating flux-pinning and pick-out depinning process from the surface to the center of the superconductor. During the process, magnetic fluxes would exist as one of states that are  flux-pinning state at volume defects and pick-out depinning state in which fluxes are moving in the superconductor. A difference of diamagnetic property occurs between pinning
state at volume defects and depinning state from the volume defects. Thus, 
diamagnetic properties of the superconductor would oscillate in $\Delta$H=$\Delta$B  region and the behavior would be observed in M-H curve. 
 The oscillation can be used for qubit by setting the pinning state at volume defects as $\ket{1}$ and the depinned state as  $\ket{0}$. This method can operate at higher temperatures than that of using  Josephson Junctions. In addition,  it is expected that the device  is quite simple and decoherences can be almost negligible. 

\end{abstract}
\pacs{74.60.-w; 74.70.Ad}

\keywords{\rm MgB$_2$, Flux pinning, (Fe, Ti) particles, Sweeping rate, Quantum bit, Oscillation behavior, Flux movement, $\Delta$H=$\Delta$B  region, Diamagnetic increase}

\maketitle
\section{Introduction} 
Quantum computers can display an outstanding performance than conventional computers for special works such as code decryption and artificial intelligence.  Their displaying speeds were expected to be rise exponentially with the number of quantum bits \cite{DiVincenzo, Bennett}. Recently, supremacy of quantum computer has been proved by several reports \cite{Arute, Zhong}. A key for broad use of quantum computer is what kind of  quantum bit is used.

There are several methods to create quantum  bits such as photon \cite{Brien, Duan}, trapped atom \cite{Blatt, Ospelkaus}, 
 nuclear magnetic resonance \cite{Jelezko, Negrevergne}, superconducting islands \cite{Nakamura, Orlando}, quantum dots and dopants in solid \cite{Koppens, Epstein}. However, these quantum bits are not easy to control in isolated state because they have decoherences by environment and are in microscopic state \cite{Ladd}. 
 Although superconducting islands, which are manipulating Josephson junction, are in mesoscopic state, they are still in difficulty because they have to work extremely low temperature owing to  decoherences by environment \cite{Mooij}.  Therefore, it is urgent to find a new platform for quantum bit that can overcome these difficulties.

 We have challenged to find a new qubit.   Recently,  an oscillation behavior was found in M-H curve (magnetic field dependence of magnetization curve) of home-made superconductor.  Here we introduce a  new platform for creating qubit, which is based on  manipulating  flux-pinning effects of a superconductor.  
  Noting things are that oscillating   energy of the behavior is high enough to be detected by current SQUID (Superconducting Quantum Interference Device) system, thus there is almost no decoherency by environment owing to its high energy. In addition, the qubit behavior is possible even 25 K when MgB$_2$ superconductor are used as a base material because MgB$_2$ showed flux-pinning effects of volume defects at 25 K \cite{Lee7, Lee9}. 
  
   If a magnetic field is applied to a superconductor containing  defects,  free energy density of the superconductor increases by H$^2/8\pi$ before fluxes penetrate into the superconductor by Meissner effect, where H is applied magnetic field before H$_{c1}$. However,  the free energy density of defects  would remain as they were because the defects are not a superconductor. Therefore, the defects which are in relatively lower free energy state than that of the superconductor would  pin quantum fluxes that are penetrated into the superconductor over H$_{c1}$, which are called flux-pinning effects of superconductor. 

If the defects in the superconductor existed in a kind of  volume defects, flux-pinning effects are more strong because of relatively larger volume comparing with planar and line defects ($\Delta G_{defect}= - \frac{(H-B)^2}{8\pi}\times\frac{4\pi}{3}r^3$, where $H$, $B$, and $r$ are applied magnetic field, magnetic induction, and radius of a volume defect, respectively). 
The pinned fluxes at volume defects would increase the max-diamagnetic property and  form $\Delta$H=$\Delta$B region on M-H curve if the number of volume defects is many, which is the region  that increased applied magnetic field ($\Delta$H) is the same as the increasing  magnetic induction ($\Delta$B) \cite{Lee4, Lee5}. In $\Delta$H=$\Delta$B region, a difference of diamagnetic properties occurs between the states that fluxes are pinned at a volume defect and they are depinned from the volume defect. 

The difference of diamagnetic property oscillates whenever fluxes would move from a volume defect to another. Therefore, a qubit can be made as a two-level system which is prepared in arbitrary superpositions of its two eigenstates, usually denoted as $\ket{1}$ and $\ket{0}$. 
In this paper, we would represent the mechanism, dynamics and experimental results for the behaviors. 

\section{Theoretical basis}

\subsection{Diamagnetic property increases by flux-pinning and  $\Delta$H=$\Delta$B  region in superconductor}
Magnetic field that have passed over lower critical field (H$_{c1}$) would  penetrate into superconductor by flux quanta.  The  penetrated fluxes  would be pinned at the first volume defect from the surface when there are volume defects in the superconductor because free energy of volume defects ($\Delta G_{defect}$) is lower than that of the superconductor. Since the pinned fluxes are arrested at the volume defect,  movement of the fluxes is blocked and diamagnetic property of the superconductor increases. 

\begin{figure}
\begin{center}
\includegraphics*[width=9cm]{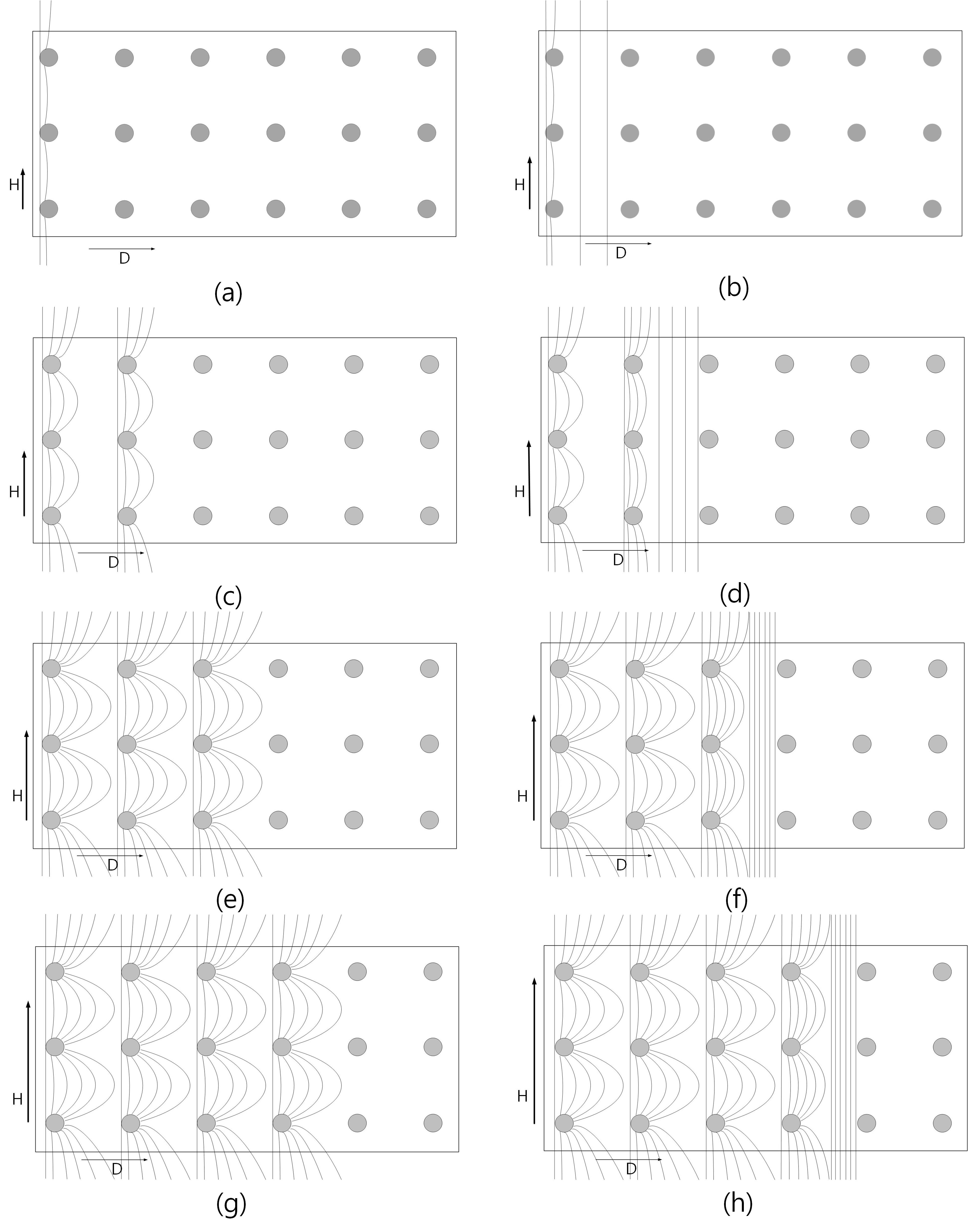}
\end{center}
\caption{Schematic representations for diamagnetic increases by flux-pinning and movements of fluxes in $\Delta$H=$\Delta$B region of first quarter of M-H curve. It is assumed that volume defects in a superconductor are arranged regularly and have the same radius. Fluxes would penetrate into the superconductor after H$_{c1}$  as magnetic field (H) increases.  
 (a): Penetrated fluxes  after H$_{c1}$ are pinned on the first volume defects from the surface. (b): The fluxes are depinned from the first volume defects and move as H increases.
  (c): Penetrated fluxes are pinned on the second volume defects from the surface. The number of pinned fluxes at a volume defect increases by increased diamagnetic property. (d): Fluxes are moving as a bundle from the second volume defects to the third. (e): Diamagnetic property increases more because of increased number of pinned fluxes at a volume defect and at last the number of pinned fluxes reaches flux-pinning limit. (f): The bundle movement of fluxes is continued as H increases. The movement of fluxes make $\Delta$H=$\Delta$B region in M-H curve if volume defects are many enough in a superconductor because of  flux-pinning limit of volume defects.  
  They show same pattern as H increases as shown (g) - (h),  which are flux-pinning and pick-out depinning process. Flux-pinning and pick-out depinning processes in other quadrants are shown in Supplementary Information.}
\label{fig1}
\end{figure}

After some fluxes have been pinned, the fluxes are pick-out depinned and move from the first volume defect by force balances which are flux-pinning force ($F_{pinning}$) and pick-out depinning force ($F_{pickout}$) \cite{Lee6}. And they are pinned again at the next volume defect which is the second volume defect from the surface.  Since two volume defects can pin fluxes at this time,  the number of fluxes that  volume defects can pin would increase, and diamagnetic property of the superconductor  increases more \cite{Lee5}. The increased diamagnetic property induce $\Delta G_{defect}$ to be deeper by B - H = 4$\pi$M, which means that a volume defect can pin fluxes more. Schematic representations for those behaviors are shown in Fig \ref{fig1}.

The number of quantum fluxes (n$^2$) pinned at a volume defect at a magnetic field (H) after H$_{c1}$ in a static state  is 
\begin{eqnarray}
n^4 - \frac{2(H-B)^2}{\alpha} r^3   = 0
\end{eqnarray} 
where  $\alpha$ is $\frac{aL'H_{c2}\Phi_o}{\sqrt{P}c}$,  $n$$^2$ is the number of quantum fluxes pinned at a spherical volume defect of radius $r$, $H_{c2}$ is upper critical field of the superconductor, $\Phi_o$ is flux quantum which is 2.07$\times10^{-7}$ G$\cdot$cm$^2$, $c$ is the velocity of light, $aL'$ is average length of quantum fluxes which are pinned and bent between volume defects (a is an average bent constant which is $1<a<1.2$ and $L'$ is the distance between volume defects in vertically packed state as shown in Fig. \ref{fig2}), 
and $P$ is the filling rate which is $\pi/4$ 
when flux quanta are pinned at a volume defect in the form of square \cite{Lee6}. 

By repeating flux-pinning and pick-out depinning process, diamagnetic property continues to increase, and at last the superconductor reaches its maximum diamagnetic property. The reason that a superconductor reaches its maximum diamagnetic property is because a volume defect has  its flux-pinning limit. The flux-pinning limit of a volume defect is related to upper critical field (H$_{c2}$) of the superconductor.  When the shortest distance between pinned fluxes at a volume defect is the same as the distance between the fluxes at H$_{c2}$, the pinned fluxes are pick-out depinned from the volume defect and   move into an inside of the superconductor. Flux-pinning effects of a superconductor could not establish if the neighborhoods of volume defect begin to be non-superconducting state. The pick-out depinning was named because the fluxes are depinned from the volume defect all at once.  When a volume defect reaches its flux-pinning limit, the diamagnetic property of the superconductor does not increase any more \cite{Lee5}. 

Flux-pinning limit of a volume defect is the maximum number of pinned fluxes at a volume defect. And the flux-pinning limit of a volume defect of radius $r$ is
 \begin{eqnarray}
n^2 = \frac{\pi r^2}{d^2}=  \frac{ r^2}{2\xi^2}
 \end{eqnarray}
 where $r$, $d$, and $\xi$ are a radius of a volume defect, the shortest distance between quantum fluxes pinned at a volume defect 
 when they have square structure, and coherence length of a superconductor, respectively  \cite{Lee4}.

When there are many volume defects in a superconductor, the superconductor forms  $\Delta$H=$\Delta$B region after reaching the maximum diamagnetic property in the M-H curve. 
The region is formed because volume defects have pinned fluxes its flux-pinning limit \cite{Lee4}. Therefore, overall magnetization is constant in the region by 4$\pi$M = B - H. The behavior of the fluxes in $\Delta$H=$\Delta$B region is related to qubit.  Figure \ref{fig1} shows  the moving behavior of pinned fluxes on volume defects from the surface to the center of the superconductor in the state that volume defects have the same radius and are distributed regularly. It is shown that pinned fluxes are moving as a bundle from a volume defect to another. The behavior of fluxes which are moving as a bundle was observed experimentally by Bonevich \cite{Bonevich}, and  $\Delta$H=$\Delta$B region is well represented in  (Fe, Ti) doped MgB$_{2}$ \cite{Lee7, Lee9}.

The pinned fluxes on volume defects would  move as a bundle due to the tension of them when they are depinned \cite{Lee6}. Therefore,  a difference in diamagnetic property occurs between  flux-pinning state on volume defects  and depinning state from the volume defects in $\Delta$H=$\Delta$B region.   
 When the fluxes are pinned at a volume defect, a diamagnetic property of the superconductor increases because the fluxes cannot move into an inside of the superconductor in the state although H increases. And when the fluxes are pick-out depinned from the volume defect, diamagnetic property of the superconductor decreases because the fluxes are moving into an inside of the superconductor. If radius of  volume defects and the distance between volume defects are set  to be constant, it would be observed that magnetization (M) of the superconductor oscillate more clearly in $\Delta$H=$\Delta$B region by resonance of pinned fluxes movement.  
 Considering the phenomenon on the basis of a oscillation, it becomes two level states with different M. Therefore, the oscillation  can be applied to a qubit. 

\section{Results and discussion}
\subsection{Calculations of flux-bundle to move at a time} 
\begin{figure}
\vspace{5cm}
\begin{center}
\includegraphics*[width=14cm]{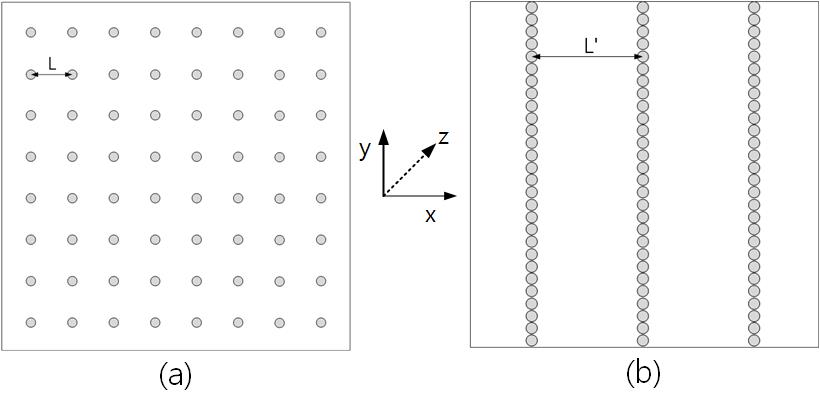}
\end{center}
\caption{ Schematic representations of regularly arranged volume defects and definition of  m$_{cps}$. It is assumed that volume defects in superconductor are regularly arranged like a lattice for calculation, which are m$^3$ volume defects in cm$^3$ superconductor. (a): If the volume defects are regularly arranged, there are  some fluxes penetrating into superconductor in unpinned state when magnetic quantum fluxes penetrate into a superconductor, which are lying parallel to x-axis and move along z-axis. (b): m$_{cps}$ is the minimum number of volume defects on which penetrating fluxes are completely pinned (2$r$$\times$m$_{cps}$=1). The volume defects are rearranged using the m$_{cps}$ concept for calculating B$_{bundle}$. In the state that volume defects are irregularly arranged, all of penetrated fluxes are  pinned on volume defects if they are many enough. }
\label{fig2}
\end{figure}

 It is assumed that all volume defects have the same radius $r$ and they are arranged regularly for resonances of pinned fluxes movement. When the number of volume defects in a superconductor is m$^3$ in unit volume (cm$^3$), the number of volume defects on a plane  is m$^2$ as shown in Fig. \ref{fig2} (a). If volume defects have a regular arrangement like lattice as shown in Fig. \ref{fig2} (a), there would be  fluxes that pass through without being pinned at volume defects when fluxes move along z-axis in the state of lying on x-axis. Thus,  a change in  arrangement  is needed for calculation of a resonated flux-bundle by pinned fluxes movement as shown in Fig. \ref{fig2} (b). Since volume defects in current experiments are arranged irregularly, all of penetrated fluxes are pinned on  volume defects if the number of volume defects is many enough. 
 
 When all of penetrated fluxes are pinned on volume defects of a plane, the corresponding minimum number of volume defects is called m$_{cps}$ (cps: closed packed state) as shown in Fig. \ref{fig2} (b).  The fluxes pinned on all  volume defects of a plane in Fig. \ref{fig2} (b)  would be pick-out depinned all at once in the state that radius of volume defects and the distance between them (L$'$) are the same, respectively.  The number of magnetic flux quanta which are pick-out depinned all at once under the state is 
  \begin{eqnarray}
B_{bundle} = n^2m_{cps}\Phi_0
 \end{eqnarray} 
where $n^2$, $m_{cps}$, and $\Phi_0$ are the number of quantum  
fluxes pinned at a volume defect,  the number of volume defects which are vertically closed packed state, 
and flux quantum, respectively \cite{Lee4}. 
 
Resonated moving fluxes  in $\Delta$H=$\Delta$B region according to a radius of volume defects is
  \begin{eqnarray}
B_{bundle} = n^2m_{cps}\Phi_0= \frac{1}{2r}\frac{ r^2}{2\xi^2}\Phi_0=\frac{r}{4\xi^2}\Phi_0 = \frac{\pi r}{2}H_{c2}
 \end{eqnarray} 
 where $m_{cps}\times{2r}$ is unit (cm) and H$_{c2}$ is $\frac{\Phi_0}{2\pi \xi^2}$.  Eq. (2) was applied because a volume defect would pin fluxes up to its flux-pinning limit in $\Delta$H=$\Delta$B region. 
Therefore, the amount of flux quanta moving simultaneously  is proportional to the radius r of  volume defects.
 If the size of a plane containing volume defects decreases, such as film, it is clear that B$_{bundle}$ also decreases proportionally.  

 In addition, diamagnetic  difference ($\Delta$M) of the oscillating behavior is as follows. When H is infinitesimally increased, the pinned fluxes are pick-out depinned from the volume defect in the state that a volume defect have pinned fluxes to its flux-pinning limit. If H is set as 0, the amplitude of the oscillation   would be  $\Delta$M=B$_{bundle}$/4$\pi$ by B – H = 4$\pi$M. 
 \begin{eqnarray} 
 \Delta M=\frac{B_{bundle}}{4\pi}=\frac{r}{16\pi \xi^2}\Phi_0 =\frac{r}{8}H_{c2}
\end{eqnarray}
 Therefore, amplitudes of the oscillations are dependent on radius of volume defects and upper critical  field of the superconductor.

Furthermore, the energy difference in an oscillation is
\begin{eqnarray} 
 \Delta G_{osc}= - \frac{(H-B)^2}{8\pi}=\frac{(4\pi\Delta M)^2}{8\pi}=2\pi\Delta M^2 = \frac{\pi}{32}r^2H_{c2}^2
\end{eqnarray}
 It is natural that $\Delta$G$_{osc}$ increase if $B_{bundle}$ increases by Eq. (5).

\subsection{Simulations of M-H curve for a superconductor}
\begin{figure}
\vspace{2cm}
\begin{center}
\includegraphics*[width=17cm]{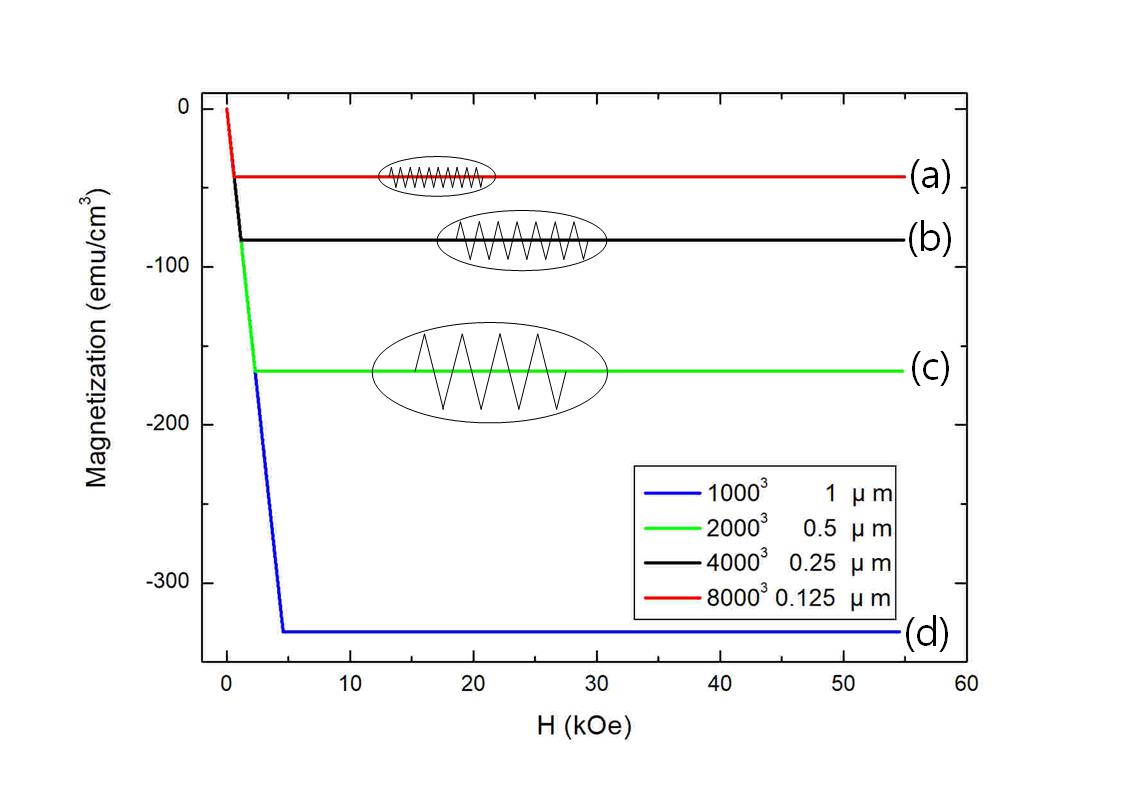}
\end{center}
\caption{Simulated M-H curves at 0 K and oscillations of magnetization M in $\Delta$H=$\Delta$B regions. When vol.\% of volume defects  is 1.56, diamagnetic properties of a  superconductor and  widths of $\Delta$H=$\Delta$B region were simulated for various conditions. It is assumed that H$_{c1}$ is 400 Oe and H$_{c2}$ is 35 T at 0 K. (a): 8000$^3$ specimen, which is the number of volume defects,   oscillates 8000 times in $\Delta$H=$\Delta$B region. Amplitude of the oscillation ($\Delta$M) was calculated to be approximately 0.55 emu/cm$^3$. (b): 4000$^3$ specimen oscillates 4000 times in $\Delta$H=$\Delta$B region. Amplitude of the oscillation ($\Delta$M) is approximately 1.1 emu/cm$^3$. (c): 2000$^3$ specimen oscillates 2000 times in $\Delta$H=$\Delta$B region. Amplitude of the oscillation ($\Delta$M) is 2.2 emu/cm$^3$. (d): It is expected that 1000$^3$ specimen would oscillate  1000 times in $\Delta$H=$\Delta$B region on the theory, but it is not shown. 
The oscillations are restricted by radius of volume defects and the distance between them because flux jump can be induced. Details for simulated M-H curves are shown in Supplementary Information.} 
\label{fig3}
\end{figure}

Figure \ref{fig3} (a) shows a simulated M-H curve in the state that  vol.\% of volume defects is 1.56  and the radius of volume defects is 125 nm, which means that the number of volume defects  is 8000$^3$ ones in cm$^3$. The M-H curve was simulated at the condition that  H$_{c1}$ and  H$_{c2}$ are 400 Oe and 35 T (Tesla) at 0 K, respectively.  Max-diamagnetic property calculated by Eq. (1) and (2), and $\Delta$H=$\Delta$B region was predicted by Eq. (3) \cite{Lee5}.  A width of $\Delta$H=$\Delta$B region (W$_{\Delta H=\Delta B}$) is $B_{bundle}\times m$ = $n^2m_{cps}m\Phi_0$, where $m$ is the number of volume defects along an axis.  
 The amplitude of the oscillation ($\Delta$M) was calculated at approximately 0.5 emu/cm$^3$.  And the $\Delta$M would oscillate 8000 times in $\Delta$H=$\Delta$B region \cite{Lee6}.

Figure \ref{fig3} (b) is a simulation that the radius of volume defects is 250 nm under the same condition with Fig. \ref{fig3} (a), which  means that the number of volume defects  is 4000$^3$ ones in cm$^3$. It is shown that max-diamagnetic property increases more owing to increase of flux-pinning limit of volume defects. 
It was calculated that the amplitude of the oscillation ($\Delta$M)  is twice of Fig. \ref{fig3} (a) because the amplitudes  have a  linear dependence of $r$ by Eq. (4).  And the $\Delta$M would oscillate 4000 times in $\Delta$H=$\Delta$B region in the state.

 A simulation  is shown in Fig. \ref{fig3} (c) when the radius of volume defects is 500 nm under the same vol.\%  with Fig. \ref{fig3} (a) and (b),  which  means that the number of volume defects  is 2000$^3$ ones in cm$^3$.  It is also shown  that max-diamagnetic property increases owing to increase of flux-pinning limit of volume defects. It is observed that $\Delta$M is twice of Fig. \ref{fig3} (b), which is approximately 2.2 emu/cm$^3$,  and the $\Delta$M would oscillate 2000 times in $\Delta$H=$\Delta$B region in the state. 

A width of $\Delta$H=$\Delta$B region is constant although max-diamagnetic properties are different according to the radius of volume defects in the state that  vol.\% of volume defects is fixed. 
 \begin{eqnarray} 
W_{\Delta H=\Delta B} = n^2m_{cps}m\Phi_0 = \frac{r^2}{2\xi^2}\frac{m}{2r}\Phi_0=\frac{1}{4\xi^2}rm\Phi_0
\end{eqnarray}
Eq. (2) and  $m_{cps}\times{2r}$=1 are applied for the equation. 
The last equation is constant because radius of volume defects decreases as the number of volume defects increases along an axis ($r\times m$ is constant).

The oscillation behavior does not happen when radius of volume defects is too large. 
An increase of radius of a volume defect means that the amount of fluxes that are moving at the same time increases when they are pick-out depinned, and the moving distance of the fluxes become longer before they are pinned again at other volume defect. If many fluxes are pick-out depinned at the same time and the moving distance became longer, those behavior would generate a lot of heat \cite{Stephen}. The heat would degrade the superconductivity of the area.  The phenomenon is called  flux jump. Therefore,  the possibility of flux jump of 1000 nm radius volume defects is relatively higher than those of the 500 nm and 250 nm radius volume defects. 
 Therefore, the oscillation behavior  for 1000 nm radius volume defects was not shown in Fig. \ref{fig3}.
However, flux jump can be hindered at the state that small volume defects are scattered around a large volume defects because moving distances of the fluxes are short.  

\subsection{ Non-uniform and irregular  distribution of volume defects, and comparison with experimental results}
Considering a state that radii of  volume defects are the same and distribution of them are irregular,  the amplitude of oscillation ($\Delta$M) will be very small because flux-pinning and pick-out depinning processes of volume defects  are continuous from a volume defect to another (B$_{bundle}$=$n^2\Phi_o$). Therefore, it is difficult to observe  an oscillation behavior in M-H curve although $\Delta$H=$\Delta$B region is formed. 

 Another state would be considered, which is the state that various radii of volume defects are present in a superconductor and the distribution of them is irregular.   As shown in Fig. \ref{fig3}, amplitudes of the oscillations  have a difference of almost 8 times according to radius of volume defects when radii of volume defects  are  from 125 nm to 1000 nm.  If  the volume defects have irregular distribution, it is natural that they are partially segregated according to kinds of radius. 

Main concerns are that oscillation behaviors 
occur in non-uniform and irregular state of volume defects, and whether the behavior can be detected or not. 
If a magnetic field is applied to the superconductor which contains volume defects in the state, penetrated flux quanta will show flux-pinning and pick-out depinning processes by each volume defect. Therefore, a difference of magnetization  by an applied magnetic field is
\begin{eqnarray} 
 \Delta M = \frac{\sum ^{ m, m, m'}_{i=1} \Delta M_{iii}}{m^2m'}
\end{eqnarray}
where $\Delta M_{iii}$, $m$, and  $m'$ are a difference of magnetization by a volume defect,  the number of volume defects along an axis (x, y), and the number of volume defects having pinned fluxes along an axis (z),
 respectively.  Therefore,  a difference of magnetization has to appear in  $\Delta$H=$\Delta$B region  of the superconductor by a segregation of volume defects if they are in non-uniform and irregular state. 

\begin{figure}
\begin{center}
\includegraphics*[width=6.7cm]{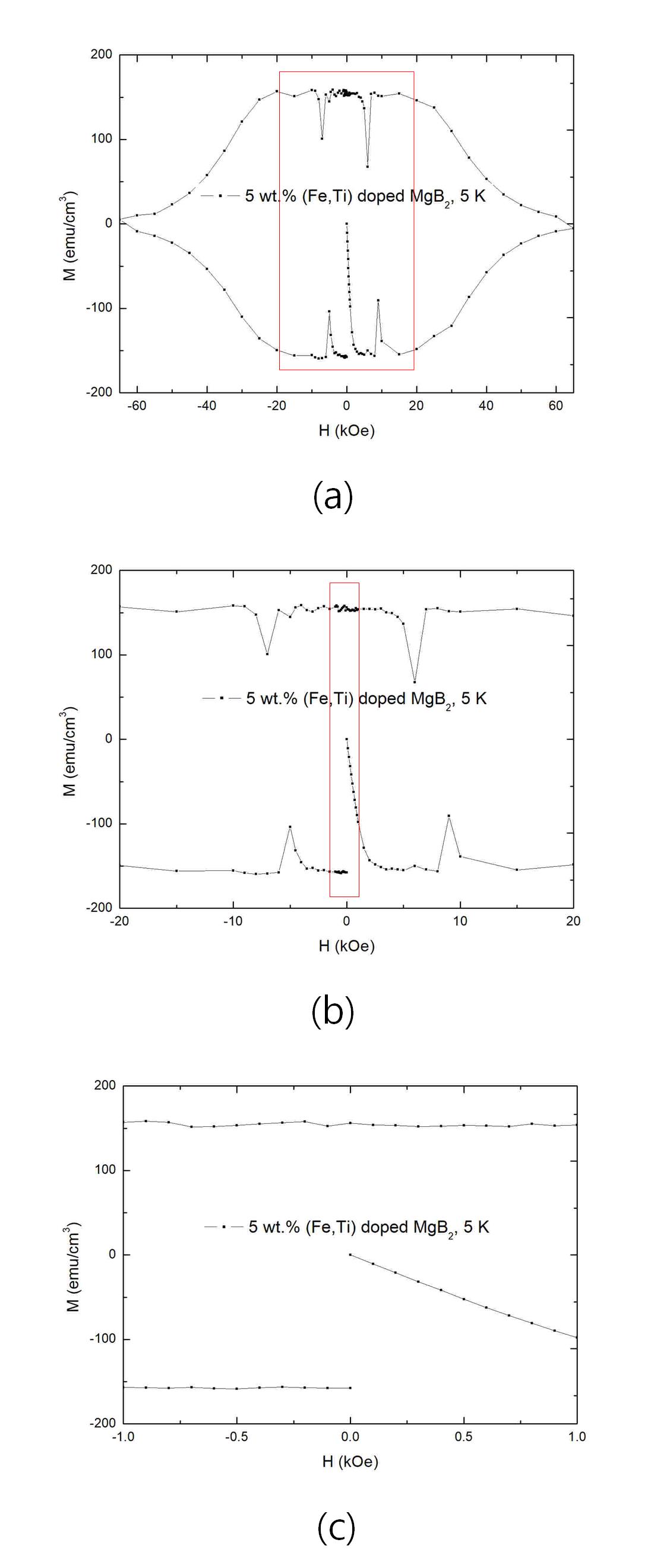}
\end{center}
\caption{M-H curve and $\Delta$H=$\Delta$B region of 5 wt.\% (Fe, Ti) doped MgB$_2$ at 5 K  in higher sweeping rate. (a): Full M-H curve.  (b):  $\Delta$H=$\Delta$B region in M-H curve, which is from -1.5 T  (Tesla) to 1.5 T. In the region, average of  magnetization (M) except flux jumps is constant and M is recovered after flux jump. Large decreases of M  are flux jumps, which are irrelevant to 
 the quantum bit behavior of fluxes. (c): Oscillation behavior of M  from  -0.1 T  to 0.1 T by pinned fluxes movement, which is most densely measured part by an applied field of 100 Oe. Oscillation behaviors are clearly observed.  Amplitudes and frequency of oscillations would be different if sweeping rates are different (see Fig. \ref{fig7}). 
 The behavior is because the number of  volume defects, which are performing flux-pinning and pick-out depinning process, are different in an applied field.  } 
\label{fig6}
\end{figure}

Figure \ref{fig6} (a) shows M-H curve of 5 wt.\% (Fe, Ti) doped MgB$_2$ at 5 K. 5 wt.\% (Fe, Ti) particles in MgB$_2$ means approximately 2 vol.\% in MgB$_2$ and average radius of (Fe, Ti) particles is 163 nm. In addition,  5 wt.\% (Fe, Ti) doped MgB$_2$  specimen contains another kind of volume defects, which are average 0.5 $\mu$m radius volume defects which are caused by low purity of boron 
 \cite{Lee4} (We reported average radius of volume defect is 1 $\mu$m, but it was confirmed that MgB$_2$ are attached on the surface of the volume defects). 
And it is certain that volume defects are in irregular state of distribution. 
Figure \ref{fig6} (b) shows oscillation behavior of $\Delta$H=$\Delta$B region, which is from -1.5 T to 1.5 T  and it shows differences of M in  $\Delta$H=$\Delta$B region  along an applied field, which includes flux jump.  Figure \ref{fig6} (c) shows a clear oscillation behavior of magnetization in  $\Delta$H=$\Delta$B region, which is the most densely measured part of the $\Delta$H=$\Delta$B regions. 
 
\begin{figure}
\begin{center}
\includegraphics*[width=6.7cm]{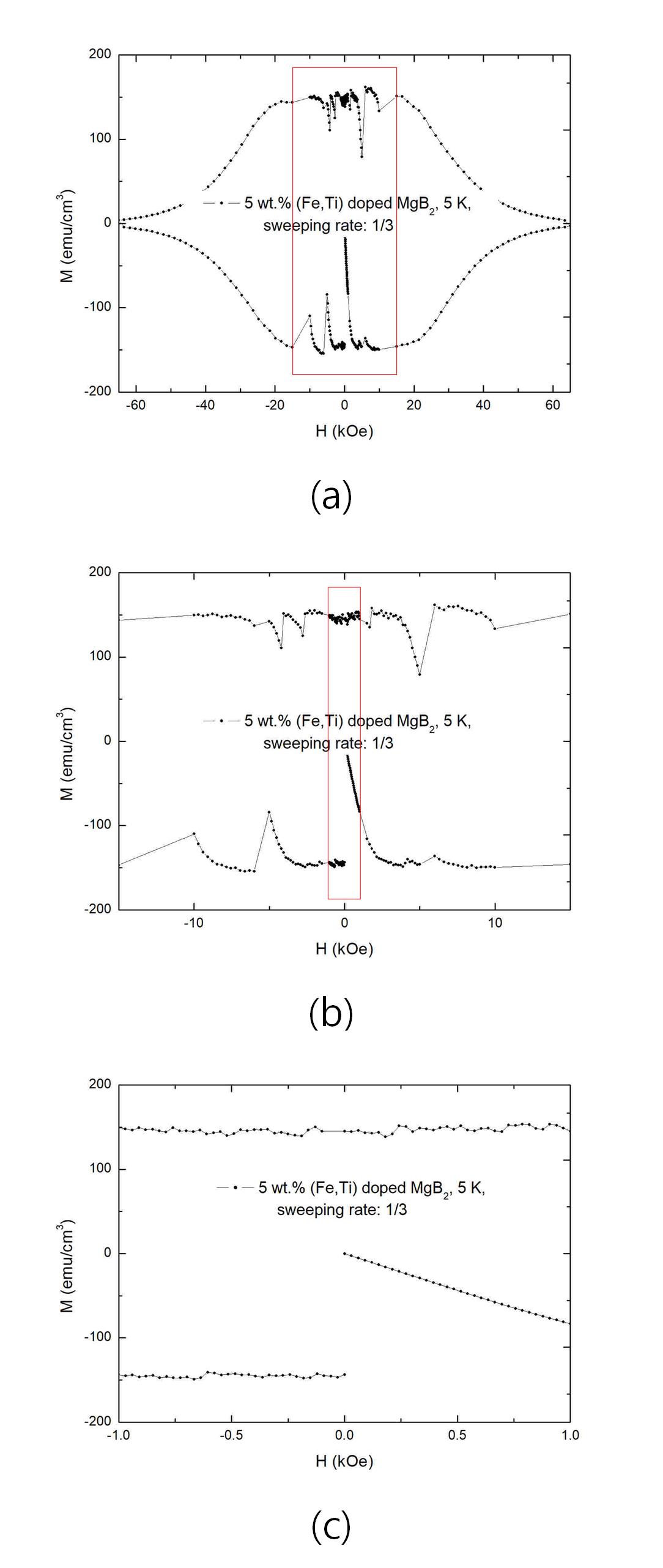}
\end{center}
\caption{M-H curve and $\Delta$H=$\Delta$B region of 5 wt.\% (Fe, Ti) doped MgB$_2$ at 5 K in lower sweeping rate which is approximately  1/3 of Fig. \ref{fig6}. The specimen is not the same as that of Fig. \ref{fig6} although compositions of the two are the same because the distribution of volume defects is different. (a): Full M-H curve. The curve is similar with  that of Fig. \ref{fig6} (a), but different in details because of different distribution of volume defects and sweeping rate.  (b): M-H curve of $\Delta$H=$\Delta$B region, which is from -1.5 T to 1.5 T. (c): Oscillation behavior of M by pinned fluxes movement from  -0.1 T to 0.1 T,  which is most densely measured part by an applied field of 30 Oe. 
It is observed that  frequency of the oscillation increased than that of Fig. \ref{fig6} (c).} 
\label{fig7}
\end{figure}

On the other hand, considering a situation that different sweeping rates are applied,  
it is predicted that a frequency of the oscillation will be changed because  the number of volume defects ($m^2m'$) affected by an applied field  are changed. 
Figure \ref{fig6} and Figure \ref{fig7} show M-H curves of 5 wt.\% (Fe, Ti) doped MgB$_2$ at 5 K, which are the sweeping rate of the former is approximately 3 times to that of the latter.
If  the sweeping rate is reduced by 1/3, 
the number of volume defects ($m^2m'$)  acting the flux-pinning and pick-out depinning process on an applied field would be reduced as many.  
Therefore, a distribution of volume defects will affect M-H curve more sensitively. Thus the oscillation behavior has to appear more frequently in the $\Delta$H=$\Delta$B  region.  

\begin{figure}
\begin{center}
\includegraphics*[width=16cm]{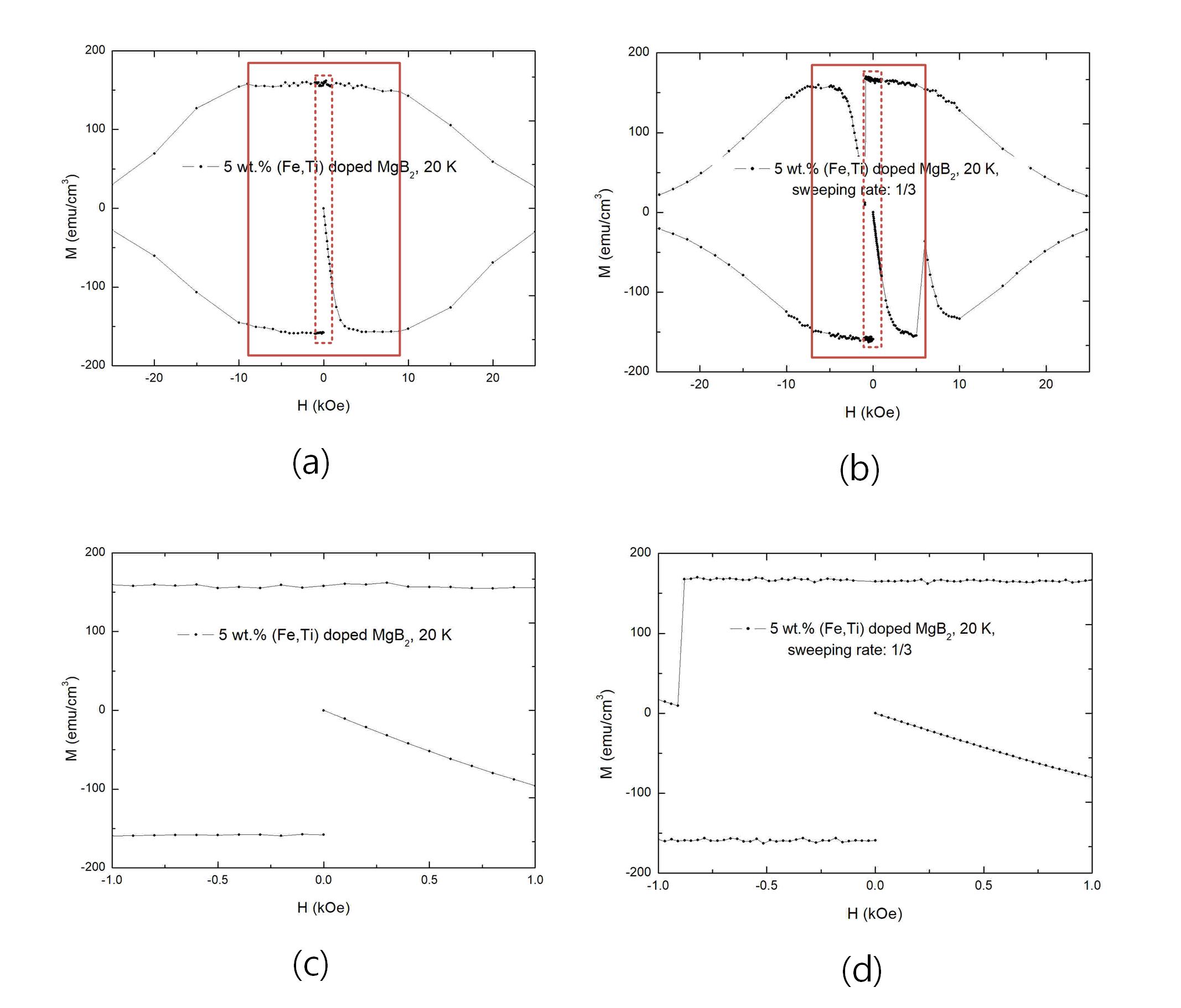}
\end{center}
\caption{M-H curves and $\Delta$H=$\Delta$B region of 5 wt.\% (Fe, Ti) doped MgB$_2$ at 20 K  in different sweeping rate. (a):  M-H curve from -2.5 T to 2.5 T  in higher sweeping rate. $\Delta$H=$\Delta$B region is determined to be from -0.8 T to 0.8 T. (b):  M-H curve from -2.5 T to 2.5 T  in lower sweeping rate, which is approximately  1/3 of (a). $\Delta$H=$\Delta$B region is determined to be from -0.65 T to 0.65 T. The specimen is not the same as that of (a) as mentioned although compositions of the two are the same because the distribution of volume defects is different. (c): Oscillation behavior of M  from  -0.1 T  to 0.1 T by pinned fluxes movement, which is most densely measured part by an applied field of 100 Oe and the inside of red dot line of (a).  (d): Oscillation behavior of M  from  -0.1 T  to 0.1 T, which is most densely measured part by an applied field of 30 Oe and the inside of red dot line of (b).
Oscillation behaviors at 10 K and 15 K are shown in Supplementary Information.} 
\label{fig10}
\end{figure}

Figure \ref{fig6} (a) and  Figure \ref{fig7} (a) show almost same magnetization behaviors, but they are different in details owing to differences of sweeping rate and and volume defects distribution. In both M-H curves, the width of  $\Delta$H=$\Delta$B region is determined to be from -1.5 T to 1.5 T. Figure \ref{fig6} (b) and  Figure \ref{fig7} (b) are  $\Delta$H=$\Delta$B region of M-H curves, which are insides of red boxes of Fig. \ref{fig6} (a) and  Fig. \ref{fig7} (a), respectively. Figures \ref{fig6} (c) and  Figure \ref{fig7} (c) show the most densely measured part of the $\Delta$H=$\Delta$B regions, which are insides of  red boxes of \ref{fig6} (b) and  Fig. \ref{fig7} (b), respectively. Comparing the two, Fig. \ref{fig7} (c) has more oscillations than those of Fig. \ref{fig6} (c). Therefore,  it  is clear that a frequency of oscillations increases  when sweeping rate is reduced in M-H curve. 
Oscillation behaviors for the same specimens at 20 K are shown  in Fig. \ref{fig10}.

Matching Eq. (8) with the two figure (Fig. \ref{fig6} (c) and Fig. \ref{fig7} (c)), $m'$ is 14.6 for 100 Oe of an applied field in Fig. \ref{fig6} (c) and  $m'$ is 4.4 for 30 Oe in Fig. \ref{fig7} (c), where $m$ is 8000 in Eq. (8). Therefore,  volume defects corresponding to 14.4 planes as shown in Fig. \ref{fig2},  performed flux-pinning and picked-out depinning process when a magnetic field was applied by 100 Oe, and average amplitudes of oscillations ($\Delta$M)  are pointed as shown in Fig. \ref{fig6} (c). And volume defects corresponding to 4.4 planes  performed the process when magnetic field are applied by 30 Oe,  average amplitudes of oscillations ($\Delta$M) are pointed as shown in Fig. \ref{fig7} (c).  
 These results support that diamagnetic increase of flux-pinning at a volume defect  and diamagnetic decreases at pick-out depinning from the volume defect, and the oscillation  behaviors of fluxes in Fig. \ref{fig3}.

\section{conclusion} 

We studied a qubit based on  flux-pinning effects of a superconductor. When  a superconductor contains volume defects many enough, magnetization (M) of the superconductor oscillates in $\Delta$H=$\Delta$B region, which provides a basis for creating quantum bit evolution. The fluxes pinned on volume defects in a superconductor always move as a bundle, qubit behavior came from the fact that diamagnetic property increases  when the fluxes are pinned at a volume defect  and diamagnetic property decreases when they are pick-out depinned from the volume defect. 
 
\section{Method}
 5 wt.\% (Fe, Ti) particle-doped MgB$_{2}$ specimens were synthesized using the nonspecial atmosphere synthesis (NAS) method \cite{Lee}. 
 The starting materials were Mg (99.9\% powder) and B (96.6\% amorphous powder) and (Fe, Ti) particles. Mixed Mg and B stoichiometry, and  (Fe, Ti) particles were added by weight. They were finely ground and pressed into 10 mm diameter pellets. (Fe, Ti) particles were ball-milled for several days, and average radius of (Fe, Ti) particles was about 0.163 $\mu$m. 
  On the other hand, an 8 m-long stainless- steel (304) tube was cut into 10 cm pieces. One side of the 10 cm-long tube was forged and welded. The pellets and excess Mg were placed in the stainless-steel tube. The pellets were annealed at 300$^o$C  for 1 hour to make them hard before inserting them into the stainless-steel tube. The other side of the stainless-steel tube was also forged. High-purity Ar gas was put into the stainless-steel tube, and which was then welded. All of the specimens had been synthesized at 920$^o$C  for 1 hour. The field dependences of magnetization were measured using a MPMS-7 which was produced by Quantum Design. During the measurement, different sweeping rates for specimens were applied for oscillation behavior of pinned fluxes movement.\\




\newpage

\end{document}